\documentclass[preprintnumbers, twocolumn,nofootinbib]{revtex4-1}

\pdfoutput=1
\usepackage{amsmath,amssymb,calc}
\usepackage{bbm}
\usepackage[pdftex]{graphicx}

\newcommand\be{\begin{equation}}
\newcommand\ee{\end{equation}}
\newcommand{\bea}{\begin{eqnarray}}
\newcommand{\eea}{\end{eqnarray}}

\def\beq{\begin{equation}}
\def\eeq{\end{equation}}
\def\id{\protect{{1 \kern-.28em {\rm l}}}}

\def\unit{\relax{\rm 1\kern-.26em I}}

\begin{document}
\title{Connection Constraints from Non-Abelian Supersymmetric Quantum Mechanics}

\author{Gianni \surname{Tallarita}}

\affiliation{Queen Mary University of London \\
Center for Research in String Theory \\
Department of Physics, \\
Mile End Road, London, E1 4NS, UK. \\ 
Email: G.Tallarita@qmul.ac.uk}

\begin{abstract}

We generalise the study of constraints imposed by supersymmetry on the Berry connection to transformations with component fields in representations of an internal symmetry group $\mathcal{G}$. Since the fields act as co-ordinates of the underlying space one finds a non-trivial extension to its structure and, correspondingly, there are new non-abelian constraints on the Berry connection. The specific case of  $\mathcal{G}=SU(2)$ is shown to constrain the connection to behave as a magnetic monopole over $\it{su}$(2), its Lie algebra.\end{abstract}

\preprint{QMUL-PH-10-05}

\maketitle

\section{Introduction}

The notion of a Berry Phase is that of a path-dependent $U(N)$ holonomy of a quantum system under adiabatic changes of external parameters \cite{Berry}\cite{Simon}. Hence, as one adiabatically varies a set of parameters $\phi_n$ around a closed path, degenerate states $\mid a\rangle$ in the quantum system undergo a holonomy
\be
U=T\exp\left(i\int_{t_i}^{t_f}A_n(\phi)\dot{\phi}^ndt \right),
\ee
where $(A_n)_{ab}=\langle b\mid\frac{\partial}{\partial\phi_n}\mid a\rangle$ is the connection over the space of parameters \cite{Zee}. Berry connections have been subject of much interest \cite{D.Tong}\cite{Vafa}\cite{Pedder}\cite{Tong2}\cite{J.N.Laia}. In particular, by considering supersymmetric quantum mechanical systems, one finds that this connection must obey specific differential equations and is thus computable even when energy eigenstates of the parameters are unknown. In \cite{D.Tong} by demanding the most general matrix-valued Lagrangian with an explicit connection term be invariant under vector multiplet supersymmetry transformations, the authors found that the connection must obey the Bogomolny Monopole equations. Similarly, when one uses chiral multiplet parameters, the connection is found to obey the $tt^*$ equations  \cite{D.Tong}\cite{Vafa}. These conditions were later generalized in \cite{J.N.Laia} where the connection was found to obey the self-dual Instanton equations, from which the previous results are obtainable via dimensional reduction. Within String theory, Berry phases have been studied arising from particular brane constructions \cite{Pedder2}\cite{Pedder3}.

In this paper we investigate the constraints on the Berry connection which result from supersymmetry transformations with component fields in representations of an internal symmetry group $\mathcal{G}$. This is a natural non-abelian extension of \cite{J.N.Laia}. In section \ref{two} we construct the most general Lagrangian, to quadratic order, built from these fields. The underlying space on which the theory lives acquires a new structure and, correspondingly, in section \ref{three} we find interesting new constraints, imposed by supersymmetry, on its connection. As a specific example the simplifying choice $\mathcal{G}=SU(2)$ is shown to constrain it to exist as a mixture of a self-dual instanton and a monopole solution over $\mathbb{R}^4\otimes \it{su}$(2). 

\section{Non-Abelian Supersymmetry and the Berry Connection}\label{two}

We are interested in applying supersymmetry to a quantum mechanical lagrangian with component fields in representations of an internal symmetry group. We focus on the case where the bosonic components of the supermultiplet act as coordinates over the space $\mathcal{S}$ on which the theory exists. In this case, the connection is simply a leading order term in the Lagrangian.\newline

In \cite{D.Tong} the bosonic fields $\phi_\mu$ act as coordinates for an underlying $\mathbb{R}^4$ manifold. When these fields are promoted to exist in representations of an internal symmetry group, this manifold acquires a new structure. The fields now parametrize
\be
\mathcal{S}=\mathbb{R}^4\otimes\mathcal{L(G)}
\ee

where $\mathcal{L(G)}$ denotes the Lie algebra of the internal group $\mathcal{G}$. This space is spanned by the bosonic coordinates $\phi_\mu^a$, with greek indices denoting the $\mathbb{R}^4$, and lower case roman alphabet indices spanning $\mathcal{L(G)}$. We will see that supersymmetry imposes non-trivial constraints on its connection. \newline

Consider the following supersymmetry transformations\footnote{We use conventions where $\sigma_\mu=(i\id_2,\sigma_i)$ and $\bar{\sigma}_\mu=(-i\id_2,\sigma_i)$}

\bea\label{transformations2}
\delta\phi^a_\mu&=&i\bar{\lambda}^a\bar{\sigma}_\mu\epsilon-i\bar{\epsilon}\sigma_\mu\lambda^a\\
\label{transformations3}
\delta\lambda^a&=&\dot{\phi}_\mu^a\bar{\sigma}^\mu\epsilon+g\epsilon(\sigma_{\mu}\bar{\sigma}_{\nu})f^{abc}\phi_b^{\mu}\phi_c^{\nu}
\eea

where $\phi^a_\mu$ are the components of a real bosonic field, $\lambda^a$ are two-component complex fermionic components, $\epsilon$ is a two-component supersymmetry generator, $f_{abc}$ are the structure constants of an internal symmetry group generated by $\chi^a$ and $g$ its corresponding coupling strength. This is the non-abelian extension of the transformations presented in \cite{J.N.Laia}. We assign engineering dimensions to these parameters consistent with the supersymmetry transformations: $[\phi]=0$, $[\psi]=\frac{1}{2}$, $[\frac{d}{dt}]=1$, $[\epsilon]=-\frac{1}{2}$ and $[g]=1$.  With these assignments we construct a supersymmetric Lagrangian with a connection term linear in time derivatives. The most general form this can take is \footnote{it is intended that only the trace of the matrix $R$ appears in this Lagrangian.}

\begin{widetext}
\bea\label{Lag}
\mathcal{L}=A_\mu^a\dot{\phi}^\mu_a+(C_{ab}\lambda^a\lambda^b+h.c)+gT_{\mu\nu}^{ab}\phi^{\mu}_a\phi^{\nu}_b+K_\mu^{ab}\bar{\lambda}_a\bar{\sigma}^\mu\lambda_b+g\phi^\mu_aE^a_\mu+gR
\eea
\end{widetext}
where $A_\mu,C,T_{\mu\nu},K_\mu$ and $R$ are functions of $\phi_\mu$, and $A^a_\mu$ is the connection. In the above $h.c$ denotes hermitian conjugation. 

\subsection{Scalar-Valued Lagrangian}

We now demand \ref{Lag} be invariant, up to a total derivative, under the transformations \ref{transformations2} and \ref{transformations3}. \newline

One finds that if, 

\bea\label{FStens}
C_{ab}&=&0\\
 F^{ab}_{\mu\nu}&=&-\frac{1}{2}\epsilon_{\mu\nu\rho\sigma}F_{\rho\sigma}^{ab}-\delta_{\mu\nu}K_{0}^{ab}\\
 E^a_\mu&=&-\partial_\mu^aR\\
 \partial^b_\nu E^a_\mu&=&-(T_{\nu\mu}^{ab}+T_{\mu\nu}^{ba})
 \eea
 
 and,
 
 \bea\label{NAres}
\partial^e_0 T^{ab}_{ij}&=&-i\epsilon_{ijk}f^{abc}K^{ce}_k\\
\partial_i^e T_{jk}^{ab}&=&i\epsilon_{ijk}f^{abc}K^{ce}_0\\
\partial^{ie}T_{ij}^{ab}&=&-if^{abc}K_j^{ce}
\eea

then the Lagrangian varies as $\delta \mathcal{L}=\dot{\Psi}$ with $\Psi=A_\mu^a\delta\phi^\mu_a$. In the above, $F_{\mu\nu}^{ab}=\partial_\mu^b A^a_\nu-\partial_\nu^a A_\mu^b$ with $\partial_\mu^a=\frac{\partial}{\partial \phi^\mu_a}$ and the indices $i,j$ indicate the $\mu=1,2,3$ components. Note that the extra structure of the underlying space means that this field strength is different to that commonly known: for example $F_{\mu\mu}^{ab}\neq0$.

\subsection{Matrix-Valued Lagrangian}

In this section we promote all functions multiplying multiplet field components in \ref{Lag} to matrices in $U(N)$.  As the functions in the Lagrangian are matrix valued, invariance under supersymmetry requires that \cite{Kentaro}

\be\label{variation}
\delta\mathcal{L}=\dot{\Theta}+i[\mathcal{L},\Theta]
\ee

where $\Theta$ is a matrix valued function of the parameters appearing in the Lagrangian and crucially the commutator here runs over the promoted algebra of the function coefficients, not over the internal symmetry group. Demanding the new Lagrangian be invariant under \ref{transformations2} and \ref{transformations3} one finds results similar to the scalar-valued case discussed previously with $\Psi=\Theta$ and the replacements of ordinary derivatives with covariant ones:

\bea\label{FStens2}
C_{ab}&=&0\\
\label{FSeq}
 F^{ab}_{\mu\nu}&=&-\frac{1}{2}\epsilon_{\mu\nu\rho\sigma}F_{\rho\sigma}^{ab}-\delta_{\mu\nu}K_{0}^{ab}\\
 E^a_\mu&=&-D_\mu^aR\\
D^b_\nu E^a_\mu&=&-(T_{\nu\mu}^{ab}+T_{\mu\nu}^{ba})
\eea
 
 and,
 
 \bea\label{NAres2}
D^e_0 T^{ab}_{ij}&=&-i\epsilon_{ijk}f^{abc}K^{ce}_k\\
\label{Releq}
D_i^e T_{jk}^{ab}&=&i\epsilon_{ijk}f^{abc}K^{ce}_0\\
D^{ie}T_{ij}^{ab}&=&-if^{abc}K_j^{ce}
\eea

where $D^a_\mu X=\partial_\mu^a X+i[A_\mu^a,X]$ and now $F_{\mu\nu}^{ab}=\partial_\mu^a A_\nu^b-\partial_\nu^b A_\mu^a+i[A_\mu^a,A_\nu^b]$. The commutators appearing here are strictly over the $U(N)$ promoted matrix structure of the Lagrangian, not over the internal group $\mathcal{G}$. \newline

 Crucially, in the limit $g\rightarrow0$, where the transformations reduce to $n$ copies of the abelian internal group and the new $g$-dependent factors in the Lagrangian disappear, one recovers the results of \cite{J.N.Laia} with $\mathcal{L(G)}=\mathbb{R}^n$. 

\section{General Constraints on the Berry Connection}\label{three}

We wish to investigate the constraints posed by supersymmetry on the connection. These constraints have a novel non-abelian contribution coming from the non-trivial internal symmetry group of the multiplet components. These contributions are most apparent from \ref{Releq} and the field strength equation \ref{FSeq}. 

Combining these in general gives an equation of the form

\be
f_{acd}F_{\mu\nu}^{ab}=-\frac{1}{2}f_{acd}\epsilon_{\mu\nu\rho\sigma}F_{\rho\sigma}^{ab}-\frac{i}{6}\delta_{\mu\nu}\epsilon_{ijk}D_i^bT_{jk}^{cd}.
\ee

Then, for $\mu \neq \nu$ we have 

\be\label{Inst}
F_{\mu\nu}^{ab}=-\frac{1}{2}\epsilon_{\mu\nu\rho\sigma}F_{\rho\sigma}^{ab}
\ee

which, at least for the case of $a=b$,  we recognise as dim$(\mathcal{G})$ copies of an instanton constraint over $\mathbb{R}^4$.  For $\mu=\nu$ one has (we have deliberately avoided to include the double index $\mu\mu$)

\be\label{NAmon}
f^{acd}F^{ab}=-\frac{i}{6}\epsilon_{ijk}D_i^bT^{cd}_{jk}
\ee

which is a novel constraint on the connection over the new structure $\mathcal{L(G)}$ of the underlying manifold. 

\subsection{The simplifying case of $\mathcal{G}=SU(2)$}
 
 General solutions of \ref{NAmon} are hard to find, however one can make a simplifying ansatz to uncover a particularly simple solution. We take the internal symmetry group to be $SU(2)$, then $f_{abc}=2i\epsilon_{abc}$ and $\mathcal{L(G)}=\it{su}$(2), so that \ref{NAmon} becomes (after contraction over a pair of indices)

\be
2\epsilon^{abc}F^{ab}=-\frac{1}{6}\epsilon_{ijk}D_{ai}T^{ac}_{jk},
\ee
which can be re-written as
\be
\frac{1}{2}\epsilon^{abc}F^{ab}=B^c
\ee
where $B^c=-\frac{1}{4}D_{1a}T_{23}^{ac}$. Together with \ref{Inst}, this means that the connection behaves as a generalisation of a $U(N)$ instanton over $\mathbb{R}^4$ and, provided $\int B^c dS_c \neq 0$ for a chosen surface $dS_c$ in $\it{su}$(2), it describes a magnetic monopole in $\it{su}$(2).

\section{Discussion}

In this paper we have shown that imposing supersymmetry with component fields in representations of an internal symmetry group $\mathcal{G}$ to the most general quantum mechanical Lagrangian built from such fields (with an explicit connection term) results in non-trivial novel constraints on the Berry connection. The underlying manifold acquires a new structure $\mathcal{L(G)}$ corresponding to the Lie algebra of the internal group. Whilst on the original $\mathbb{R}^4$ the connection is always constrained to obey the self-dual instanton equations, it is on this new structure that the new features are observed.  In the simplest case, where $\mathcal{L(G)}=\mathbb{R}^n$ and the chosen group is abelian one recovers the results of \cite{J.N.Laia}. Furthermore, in the simplifying case of $\mathcal{G}=SU(2)$ the new constraints are shown to be those of a monopole over $\it{su}$(2). In general,  \ref{NAmon} is a novel constraint on the Berry connection over $\mathcal{L(G)}$. It would be interesting to investigate whether a different choice for $\mathcal{G}$ also gives a known solution for the Berry Connection. We leave this for further work. 

In \cite{Konyushikhin} and \cite{Ivanov}, similar constraints on an $SU(2)$ connection were found by a harmonic superspace approach. An explicit form for the superfield action was given for the case where the underlying manifold is $\mathbb{R}^4$. In our case, where the manifold becomes $\mathbb{R}^4\otimes \mathcal{L(G)}$, we expect a similar argument to hold, even though no explicit action was given here. This has interesting connections to string theory, from which this general construction is thought to exist in a low-dimensional limit or a particular brane construction.\\

\begin{acknowledgments}
The author would like to thank D.Tong and D.R.Gomez for providing essential insight into this work and is also grateful to D.S.Berman and D.Thompson for useful discussions. GT is supported by an EPSRC grant.
\end{acknowledgments}


\begin{thebibliography}{1}


\bibitem{Berry}
M.V.Berry,``Quantal phase factors accompanying adiabatic changes", Proc. Roy. Soc. Lond, A392:45-57,1984.

\bibitem{Simon}
B. Simon,``Holonomy, the quantum adiabatic theorem, and BerryÕs phase'', Phys. Rev. Lett. 51, 2167 (1983)

\bibitem{Zee}
F. Wilczek and A. Zee,``Appearance Of Gauge Structure In Simple Dynamical Systems'', Phys. Rev. Lett. 52, 2111 (1984)

\bibitem{D.Tong}
D.Tong, J.Sonner,``Berry Phase and Supersymmetry", JHEP 0901, 063 (2009) arXiv:0810.1280v3 [hep-th].

\bibitem{Vafa}
S.Cecotti and C.Vafa,``Topological antitopological fusion", Nucl.Phys., B367:359-461,1991

\bibitem{Pedder}
C.Pedder, J.Sonner and D.Tong, ``The Geometric Phase in Supersymmetric Quantum Mechanics", Phys. Rev. D 77 (2008) 025009, arXiv:0709.0731 [hep-th].

\bibitem{Tong2}
J.Sonner and D.Tong,``Non-Abelian Berry Phases and BPS Monopoles", Phys.Rev.Lett.102:191801,2009 arXiv:0809.3783 [hep-th].

\bibitem{J.N.Laia}
J.N.Laia``Non-Abelian Berry phase, Instantons and N=(0,4) Supersymmetry", arXiv:1003.4751 [hep-th].


\bibitem{Pedder2}
C. Pedder, J. Sonner and D. Tong,``The Geometric Phase and Gravitational Precession
of D-BranesÓ, Phys. Rev. D 76, 126014 (2007) arXiv:0709.2136 [hep-th].

\bibitem{Pedder3}
C. Pedder, J. Sonner and D. Tong,``The Berry Phase of D0-BranesÓ, JHEP 0803, 065
(2008) arXiv:0801.1813 [hep-th].


\bibitem{Kentaro}
M.Herbst, K.Hori and D.Page,``Phases Of N=2 Theories In 1+1 Dimensions With Boundary", arXiv:0803.2045 [hep-th].

\bibitem{Konyushikhin}
E.Ivanov, M.Konyushikhin and A.Smilga ``SQM with Non-Abelian Self-Dual Fields: Harmonic Superspace Description'', JHEP 1005, 033 (2010) arXiv:0912.3289 [hep-th]

\bibitem{Ivanov}
E.Ivanov and M.Konyushikhin,``N=4, 3D Supersymmetric Quantum Mechanics in Non-Abelian Monopole Background", arXiv:1004.4597v1 [hep-th].

\end{thebibliography}
\end{document}